# Reply to "Issues arising from benchmarking single-cell RNA sequencing imputation methods"

Mo Huang[1] & Nancy R. Zhang[1]

1) Department of Statistics, The Wharton School, University of Pennsylvania, Philadelphia, PA
Correspondence:
Nancy R. Zhang
nzh@wharton.upenn.edu

In our Brief Communication[1], we presented the method SAVER for recovering true gene expression levels in noisy single cell RNA sequencing data. We evaluated the performance of SAVER, along with comparable methods MAGIC[2] and scImpute[3], in an RNA FISH validation experiment and a data downsampling experiment. In a Comment [arXiv:1908.07084v1][4], Li & Li were concerned with the use of the downsampled datasets, specifically focusing on clustering results obtained from the Zeisel *et al.*[5] data. Here, we will address these comments and, furthermore, amend the data downsampling experiment to demonstrate that the findings from the data downsampling experiment in our Brief Communication are valid.

Briefly, the downsampling experiment involved four diverse publicly available single cell datasets. For each of the datasets, we selected high quality cells and highly expressed genes to mimic true expression levels since the measurements for these cells and genes are less affected by noise. The resulting expression matrix serves as the reference dataset, a stand-in for the true expression $\lambda_{gc}$. We then sampled each expression value from a Poisson noise model, which has been shown to accurately describe the technical noise in scRNA-seq experiments with UMIs[6,7]. Specifically, the observed number of UMIs $Y_{gc}$ can be modeled as $Y_{gc} \sim Poisson(\tau_c \lambda_{gc})$, where $\tau_c$ is the cell-specific sequencing efficiency and $\lambda_{gc}$ is the true expression of the reference dataset. $\tau_c$ was simulated from a Gamma distribution with mean efficiency of 5% or 10% depending on the dataset, which has been shown to be appropriate[8].

Li & Li incorrectly state that the downsampled datasets were simulated from the Poisson-Gamma model used in the SAVER method. The SAVER method assumes a prior Gamma distribution on $\lambda_{gc}$ which SAVER tries to estimate. This is different from our use of a Gamma distribution as a convenient way to describe the variation of the sequencing efficiency $\tau_c$.

Li & Li claim that the downsampled datasets are not representative of real scRNA-seq data by comparing the mean, standard deviation, and zero fraction of the downsampled dataset with the original Zeisel *et al.* dataset. We agree that even though we tried to match the mean expression and total zero fraction of the downsampled datasets with the original datasets, the distributions of these characteristics differ from the original datasets. However, Li & Li do not investigate the effect of this deviation on downstream analysis.

Li & Li also reported that the average coefficient of determination between a gene's expression in the downsampled dataset and its expression in the original dataset is only 14%. This, however, is the intended consequence of the downsampling experiment — downsampling simulates noise in the data generation process, so that the downsampled data is a noisy representation of the reference data, which is a subset of the original data. Our goal is to evaluate how accurately each method can recover the reference data from the noisy downsampled data.

In light of these comments, we decided to amend the downsampling procedure from the reference dataset to accurately represent the original data. To do this, we randomly selected a subset of genes and cells from the original dataset, calculated the mean expression and library size, and downsampled from the reference data to match the mean expression and library size selected from the original data. Figure 1 displays the mean expression, standard deviation, and zero fraction for genes in the original Zeisel *et al.* dataset and the new downsampled dataset, which can be compared directly to Figure 1 in Li & Li's Comment. The characteristics of the new downsampled dataset is almost identical to those of the original dataset.

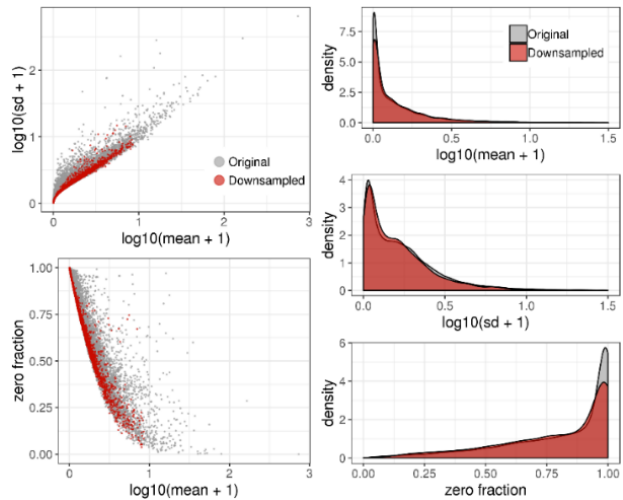

**Figure 1** Scatterplots and density plots comparing mean expression, standard deviation, and zero fraction for genes in the original Zeisel *et al.* dataset and the new downsampled dataset.

Using this amended downsampling technique, we constructed new downsampled datasets for the Baron *et al.*[9], Chen *et al.*[10], La Manno *et al.*[11], and Zeisel *et al.*[5] datasets. We then ran

SAVER v1.0.0, MAGIC v1.0.0, and scImpute v0.0.3 on these downsampled datasets and analyzed their correlation with the reference, correlation matrix distance, and clustering as we had done in Figure 2 of our Brief Communication (Figure 2). The results obtained on the new downsampled datasets are in agreement with the findings in our Brief Communication.

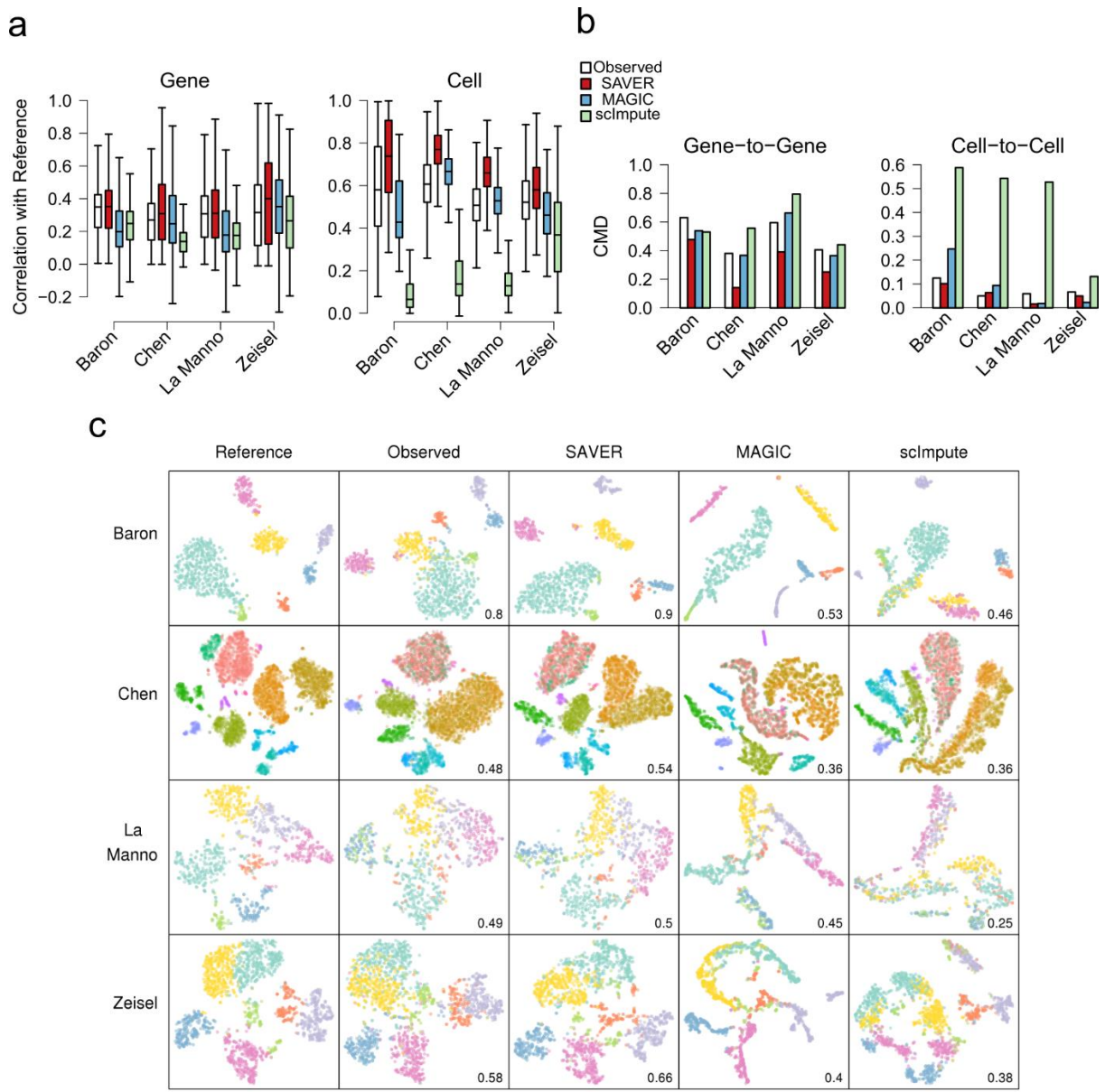

**Figure 2** (**a**) Performance of algorithms measured by correlation with reference data, on the gene level (left) and on the cell level (right). (**b**) Comparison of gene-to-gene and cell-to-cell correlation matrices of recovered values with the true correlation matrices. (**c**) Cell clustering and t-SNE visualization of data from Baron *et al.*, Chen *et al.*, La Manno *et al.*, and Zeisel *et al*. The Jaccard index of the downsampled observed dataset and recovery methods as compared with the reference classification is shown.

In their Comment, Li & Li also argue that real data should be used to evaluate imputation methods. As a result, they applied SAVER, MAGIC, and scImpute to the full Zeisel *et al.* dataset and performed hierarchical clustering to re-identify cell types already identified by Zeisel *et al.* They showed that scImpute obtained comparable or higher clustering accuracy than SAVER when compared with the Zeisel *et al.* labels. We believe that there is a flaw to this logic — imputation and expression recovery is meant to ideally improve upon the findings obtained from the original data and to uncover novel relationships that were obscured by noise. Thus, comparison to findings with the original data does not tell us how well the imputation/recovery method works.

Furthermore, the choice of clustering method is extremely important in evaluating performance. Li & Li performed hierarchical clustering on the first ten principal components, which is not a standard technique in identifying cell types. In fact, Li & Li's Figure 1c shows that using hierarchical clustering leads to extremely poor clustering metrics on the raw Zeisel *et al.* data, from which the true labels were originally generated. For example, an adjusted Rand index (ARI) of 0 indicates that two sets of labels overlap by random chance only, whereas an ARI of 1 indicates a perfect overlap. In the boxplot of the 9 clusters, the ARI between the original Zeisel *et al.* labels and the labels obtained through hierarchical clustering is only slightly above 0. Such a large discrepancy suggests hierarchical clustering is an unreliable method to identify cell types, and thus it is unclear how one should interpret comparisons based on this method.

It is also important to note that the number of major cell types identified by Zeisel *et al.* was provided as input to scImpute as the number of clusters in the data, while the other imputation/recovery methods were blind to this crucial information, and so Li & Li's re-analysis of the full Zeisel data unrealistically favors scImpute. It is rare for scientists to know the number of clusters prior to analyzing the data, and so this requirement by scImpute does not reflect typical analyses of scRNA-seq data by the scientific community.

We strived to make our experiments as comparable to real data analysis scenarios as possible. This is the reason why we decided to use the popular single cell analysis package Seurat[12] to perform clustering, to use RNA FISH as a separate source of evaluation, and to use the clusters obtained from the reference data to compare with the clusters obtained from the downsampled data. The downsampling scheme is preferable to generating completely synthetic datasets where gene-gene interactions and biological variation cannot be accurately captured. We acknowledge Li & Li's concern with the use of synthetic datasets in method evaluation but we believe that third party evaluations[13–15] and feedback from the scientific community are the ultimate source of benchmarking.